\documentclass[article,preprint,superscriptaddress]{revtex4-1}
\usepackage{graphicx}
\usepackage{dcolumn}
\usepackage{amssymb}
\usepackage{amsmath}
\usepackage{bm}
\usepackage{pifont}
\usepackage{epsfig}
\usepackage{psfrag}
\usepackage[usenames]{color}
\begin{document}
\title{Direct measurements of growing amorphous order and non-monotonic dynamic correlations in a colloidal glass-former}
\author{K. Hima Nagamanasa$^{\ast}$}
\affiliation{Chemistry and Physics of Materials Unit, Jawaharlal Nehru Centre for Advanced Scientific Research, Jakkur, Bangalore - 560064, INDIA}
\author{Shreyas Gokhale}
\affiliation{Department of Physics, Indian Institute of Science, Bangalore - 560012, INDIA}
\author{A. K. Sood}
\affiliation{Department of Physics, Indian Institute of Science, Bangalore - 560012, INDIA}
\affiliation{International Centre for Materials Science, Jawaharlal Nehru Centre for Advanced Scientific Research, Jakkur, Bangalore - 560064, INDIA}
\author{Rajesh Ganapathy}
\affiliation{International Centre for Materials Science, Jawaharlal Nehru Centre for Advanced Scientific Research, Jakkur, Bangalore - 560064, INDIA}
\date{\today}
\draft
\maketitle
\renewcommand{\thefootnote}

While the transformation of flowing liquids into rigid glasses is omnipresent, a complete understanding of vitrification remains elusive. Of the numerous approaches \cite{berthier2011theoretical,kirkpatrick1989scaling,chandler2009dynamics,tarjus2005frustration} aimed at solving the glass transition problem, the Random First-Order Theory (RFOT) \cite{kirkpatrick1989scaling,lubchenko2007theory} is the most prominent. However, the existence of the underlying thermodynamic phase transition envisioned by RFOT remains debatable, since its key microscopic predictions concerning the growth of amorphous order and the nature of dynamic correlations lack experimental verification. Here, by using holographic optical tweezers, we freeze a wall of particles in an \textit{equilibrium configuration} of a 2D colloidal glass-forming liquid and provide direct evidence for growing amorphous order in the form of a static point-to-set length. Most remarkably, we uncover the non-monotonic dependence of dynamic correlations on area fraction and show that this non-monotonicity follows directly from the change in morphology of cooperatively rearranging regions, as predicted by RFOT \cite{stevenson2006shapes}. Our findings suggest that the glass transition has a thermodynamic origin.

In a seminal paper dated nearly fifty years ago, Adam and Gibbs \cite{adam1965temperature} associated the rapid growth of the relaxation time with a decrease in the liquid's configurational entropy $S_c$. Within RFOT, $S_c$ is related to the number of metastable minima in the potential energy landscape of the liquid that can be explored by the system at a given temperature. RFOT further claims that the supercooled liquid freezes into a mosaic whose domains correspond to configurations in these metastable minima \cite{berthier2011theoretical}. The typical domain size is inversely related to $S_c$ and diverges at the `ideal' glass transition temperature, where $S_c$ vanishes. The existence of a growing static `mosaic' length scale that serves as a clear indicator of the glass transition is therefore intrinsic to RFOT \cite{kirkpatrick1989scaling}, although a systematic procedure for measuring it from point-to-set correlations was established much later \cite{bouchaud2004on}. Since the findings of \cite{bouchaud2004on}, a variety of growing static length scales have been identified and computed in numerical simulations \cite{biroli2008thermodynamic,tanaka2010critical,kurchan2011order,karmakar2012direct}. Of these, the point-to-set correlation length $\xi_{PTS}$ \cite{biroli2008thermodynamic} is of central importance in the RFOT scenario, as it follows directly from the mosaic picture. $\xi_{PTS}$ is measured by freezing a subset of particles in the liquid's equilibrium configuration, and examining their influence on the configuration of the remaining free particles. As such, $\xi_{PTS}$ quantifies the spatial extent of the influence of amorphous boundary conditions, thereby providing an estimate of the typical domain size of the mosaic. In addition to its close connection to RFOT, it has been shown analytically, that a divergence in the relaxation time is indeed associated with a diverging $\xi_{PTS}$ \cite{montanari2006rigorous}. $\xi_{PTS}$ was first extracted in simulations by pinning all particles outside a spherical cavity and examining the configurations of free particles inside the cavity \cite{biroli2008thermodynamic}. Subsequently, the growth of $\xi_{PTS}$ has been studied for various pinning geometries \cite{berthier2012static} as well as in various simulated glass-formers \cite{hocky2012growing}. Of particular importance is the case in which the pinned particles form a single amorphous wall. Using this geometry, recent simulations \cite{kob2012non} have discovered that in addition to $\xi_{PTS}$, one can define a dynamic correlation length $\xi_{dyn}$ that evolves non-monotonically with temperature across the mode coupling crossover. It was surmised that this non-monotonicity reflects a change in the morphology of cooperatively rearranging regions (CRRs), which according to RFOT are string-like at high temperatures and compact close to the glass transition \cite{stevenson2006shapes}. However, the crucial and long-standing microscopic predictions of RFOT pertaining to growing point-to-set correlations and the morphology of CRRs remain untested in experiments. Point-to-set correlations cannot be investigated in atomic and molecular glass-formers, since the dynamics of their constituent particles cannot be traced, and it is not possible to freeze a subset of particles in an equilibrium configuration. These problems can be alleviated in colloidal glass-formers, and in fact, the random pinning geometry was realized in a very recent experiment \cite{gokhale2014growing}. Given this advance in colloid experiments, testing the key predictions of RFOT directly in colloidal glass-formers would constitute a major step in unravelling the nature of the glass transition.
 
\begin{figure*}
  \centering
  \includegraphics[height=14cm]{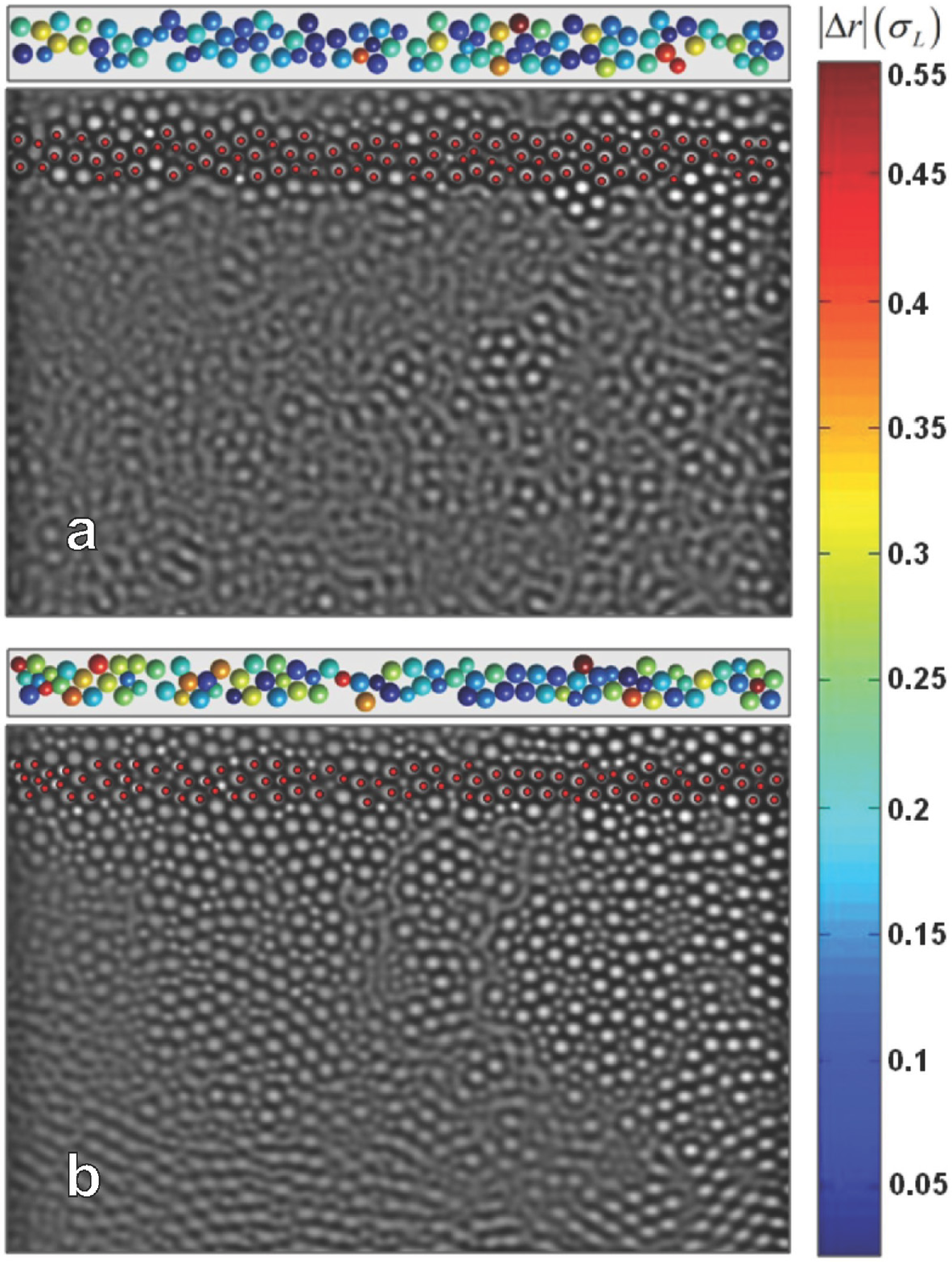}
  \caption{\textbf{Visualization of the amorphous wall.} The underlying grey scale images have been generated by time-averaging snapshots over 30$\tau_{\alpha}$ for $\phi = 0.68$ (a) and $\phi = 0.76$ (b), respectively. The red circles correspond to the coordinates of the trapped particles that form the amorphous wall. The spheres at the top of the images in (a-b) constitute the pattern whose fast Fourier Transform was fed into the spatial light modulator (SLM). Spheres are colour coded according to the distance between the initial and time-averaged particle positions in units of $\sigma_L$.}
  \label{Fig1}
\end{figure*}

We performed optical video microscopy experiments on a binary mixture of small and large polystyrene colloids of diameters $\sigma_{S}$ and $\sigma_{L}$, respectively (see Methods for experimental details). As mentioned before, measuring point-to-set correlations requires pinning particles in an equilibrium configuration of the liquid, which is experimentally challenging. In colloidal systems, this can be realized by manipulating light fields using holographic optical tweezers \cite{gokhale2014growing,irvine2013dislocation}. Here, we demonstrate the power of this technique by pinning an amorphous wall of particles in a 2D colloidal glass-forming liquid. We first captured a bright field image of the sample and extracted particle coordinates within a strip of width $\sim$ 2$\sigma_{L}$ along the longer dimension of the field of view. We then calculated the hologram and fed it to a spatial light modulator (SLM) (Boulder Nonlinear Systems, Inc.), which in turn created traps at the desired positions. Further, the use of a SLM ensured that all the particles constituting the wall were frozen simultaneously (see Supplementary Video S1). To ensure that the particles thus pinned are indeed a part of the liquid's equilibrium configuration, we superimposed the coordinates of these particles on time averaged images of the sample in the presence of the amorphous wall, for two different area fractions $\phi$ (Fig. \ref{Fig1}). Particles forming the wall appear bright in the time-averaged images owing to their negligible mobility, and can be easily identified. We observe that by and large the initial set of particle coordinates and the centres of pinned particles acquired from the time averaged images are separated by a distance smaller than the cage size, which shows that almost all the particles forming the amorphous wall are pinned in an equilibrium configuration of the colloidal liquid. Interestingly, from the time averaged images in Fig. \ref{Fig1}, we see that clusters of immobile particles extend further away from the wall for $\phi=0.76$ as compared to $\phi=0.68$, suggesting that the influence of the wall is felt over longer distances with increasing area fraction.

\begin{figure*}
  \centering
  \includegraphics[width=\textwidth]{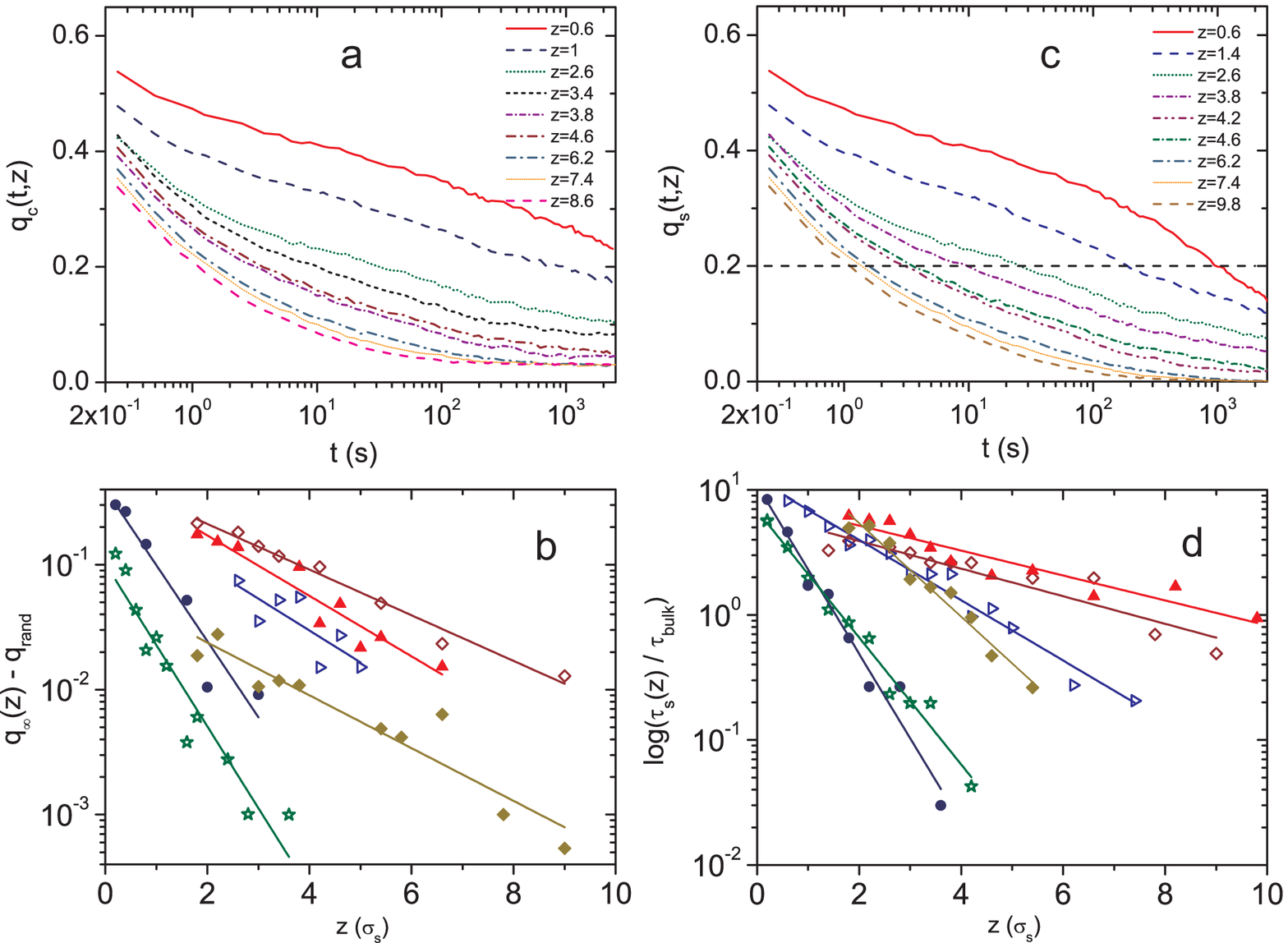}
  \caption{\textbf{Overlap functions and relaxation times.} (a) The total overlap $q_{c}(t,z)$ for $\phi = 0.74$. Different colours represent different $z$. (b) $q_{\infty}(z) - q_{rand}$ versus z for $\phi =$ 0.68 ({\color{Blue} $\boldsymbol \bullet$}), $\phi =$ 0.71 ({\color{Green} $\boldsymbol \star$}), $\phi =$ 0.74 ({\color{Blue} $\boldsymbol \triangleright$}), $\phi =$ 0.75 ({\color[rgb]{0.71,0.65,0.26} $\boldsymbol \blacklozenge$}), $\phi =$ 0.76 ({\color{Red} $\boldsymbol \blacktriangle$}) and $\phi =$ 0.79 ({\color{Brown} $\boldsymbol \diamond$}). (c) The self overlap $q_{s}(t,z)$ for $\phi = 0.74$. Different colours correspond to different $z$. (d) $\text{log}(\tau_s(z) /  \tau_s^{bulk})$ as a function of z. The colors and symbols in (\textbf{d}) are identical to those in (\textbf{b}). In (b) and (d), the solid lines are exponential fits of the forms given in Eqns \ref{XiPTS} and \ref{xidyn}, respectively.}
  \label{Fig2}
\end{figure*}

To extract $\xi_{PTS}$ and $\xi_{dyn}$ we followed the protocol described in \cite{kob2012non}. We first extracted $\xi_{PTS}$ by calculating the total overlap function, $q_c(t,z)$, at various distances $z$ from the pinned wall. To this end we divided the field of view into boxes of size $0.25\sigma_{S}$, and computed $q_c(t,z)$ for all boxes that lie at a fixed distance away from the wall, using the equation
\begin{equation}
q_c(t,z) = \frac{\sum_{i(z)}\langle n_i(t)n_i(0)\rangle}{\sum_{i(z)} \langle n_i(0)\rangle}
\end{equation} 
where $i$ is the box index, $n_{i}(t) = 1$ if the box contains a particle at time $t$ and $n_{i}(t) = 0$ otherwise. Fig. \ref{Fig2}a shows $q_c(t,z)$ at various $z$ for $\phi = 0.74$. The box size $0.25\sigma_{S}$ was chosen to be larger than the cage size, $\sim 0.14\sigma_{S}$ to avoid spurious overlap fluctuations due to cage rattling, but small enough to provide sufficient statistics to compute $\xi_{PTS}$ and $\xi_{dyn}$. By definition, $q_c(t,z)$ measures the overlap between configurations at two different times at a given distance from the wall. Since $q_c(t,z)$ is insensitive to particle exchanges, in the limit of long times and large distances from the wall, it attains a finite asymptotic bulk value $q_{rand}= q_c(t\rightarrow\infty,z\rightarrow\infty)$ corresponding to the probability of occupation of a box. Consistent with simulations \cite{kob2012non}, we observe that the presence of the wall influences the asymptotic value of $q_c(t\rightarrow\infty,z) = q_{\infty}(z)$, such that $q_{\infty}(z) > q_{rand}$ (Fig. \ref{Fig2}a). Further, as expected, $q_{\infty}(z)$ decreases with $z$ in the vicinity of the wall. This is also evident from (Fig. \ref{Fig2}a), where the $q_c(t,z)$ profiles for large $z$ overlap almost completely. We observe that $q_{\infty}(z)$ - $q_{rand}$ decays exponentially with $z$ (Fig. \ref{Fig2}b), which allowed us to extract $\xi_{PTS}$ from the relation 
\begin{equation}
q_{\infty}(z) - q_{rand} = B\:\text{exp}(-z/\xi_{PTS})
  \label{XiPTS}
\end{equation} 

Having computed $\xi_{PTS}$, we computed $\xi_{dyn}$ from the self part of the overlap function, $q_{s}(t,z)$: 
\begin{equation}
q_s(t,z) = \frac{\sum_{i(z)}\langle n_{i}^{s}(t)n_{i}^{s}(0)\rangle}{\sum_{i(z)} \langle n_{i}^{s}(0)\rangle}
\end{equation} 
where once again, $i$ is the box index, and $n_{i}^{s}(t) = 1$ if the box is occupied by the \textit{same} particle at time $t$ and $n_{i}^{s} = 0$ otherwise. $q_s(t,z)$ is similar to the self-intermediate scattering function calculated for the wave vector corresponding to the box size. Unlike $q_c(t,z)$, $q_s(t,z)$ is sensitive to particle exchanges and reaches zero at long times, when all the particles undergo a displacement larger than the box size. Owing to its similarity with the self-intermediate scattering function, $q_s(t,z)$ yields relaxation times $\tau_s(z)$ at different distances $z$ from the wall \cite{kob2012non}. Due to the limited temporal resolution in our experiments, we defined $\tau_s(z)$ as the time taken for $q_{s}(t,z)$ to decay to 0.2 instead of $1/e$ \cite{starrjcp}. Fig. \ref{Fig2}c shows $q_{s}(t,z)$ at various $z$ for $\phi = 0.74$. As expected, $\tau_s(z)$ approaches its bulk value $\tau_s^{bulk}$ for large $z$. In accordance with simulations \cite{kob2012non}, we find that the dynamic length scale $\xi_{dyn}$ (Fig. \ref{Fig2}d) can be extracted from the equation 
\begin{equation}
\text{log}(\tau_s(z)) = \text{log}(\tau_s^{bulk}) + B_s \text{exp}(-z/\xi_{dyn})
  \label{xidyn}
\end{equation}
 
\begin{figure}
  \centering
  \includegraphics[height=17cm]{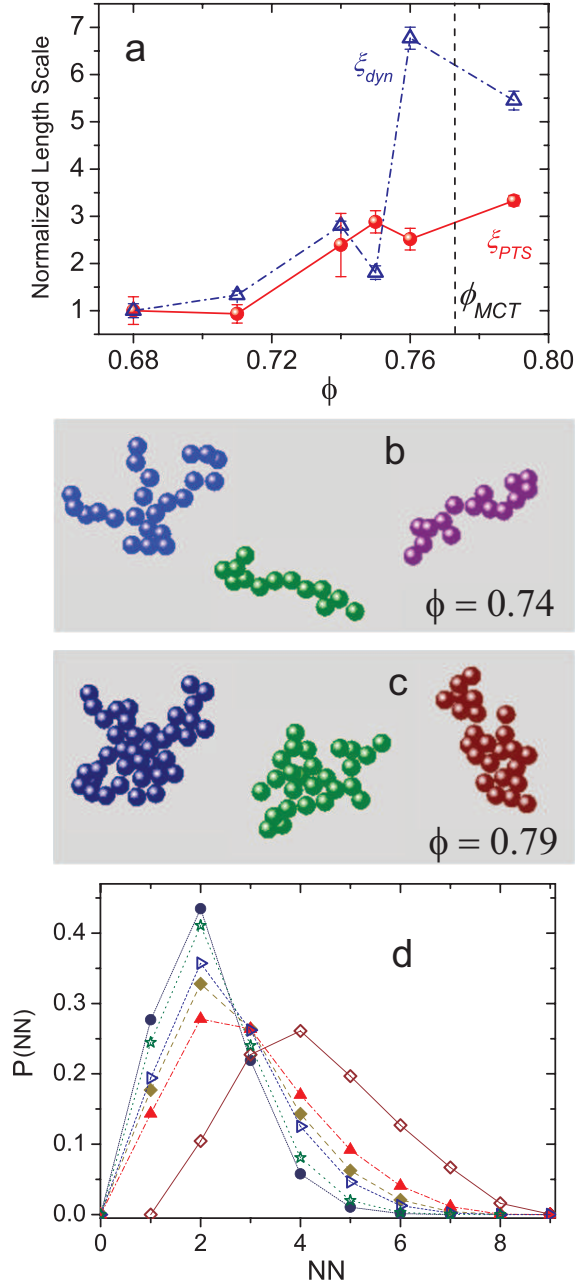}
  \caption{\textbf{Static and dynamic length scales and morphology of CRRs.} (a) Point-to-set length scale $\xi_{PTS}$ ({\color{Red} $\boldsymbol \bullet$}) and dynamic length scale $\xi_{dyn}$ ({\color{Blue} $\boldsymbol \triangle$}) normalized by their respective values at $\phi =$ 0.68. The error bars have been obtained from the exponential fits. The dotted black line indicates the mode coupling crossover $\phi_{MCT}$. (b-c) Representative cluster morphologies for $\phi =$ 0.74 and $\phi =$ 0.79 respectively. (d) Probability distribution of the number of mobile nearest neighbours, $P(NN)$, for $\phi =$ 0.68 ({\color{Blue} $\boldsymbol \bullet$}), $\phi =$ 0.71 ({\color{Green} $\boldsymbol \star$}), $\phi =$ 0.74 ({\color{Blue} $\boldsymbol \triangleright$}), $\phi =$ 0.75 ({\color[rgb]{0.71,0.65,0.26} $\boldsymbol \blacklozenge$}), $\phi =$ 0.76 ({\color{Red} $\boldsymbol \blacktriangle$}) and $\phi =$ 0.79 ({\color{Brown} $\boldsymbol \diamond$}).}
  \label{Fig3}
\end{figure}

Having extracted $\xi_{PTS}$ and $\xi_{dyn}$ from overlap functions, we studied the variation of these length scales with the area fraction $\phi$ on approaching the glass transition Fig. \ref{Fig2}b \& d. We find that in concord with simulations \cite{kob2012non}, $\xi_{PTS}$ grows monotonically with $\phi$ (Fig. \ref{Fig3}a). This finding constitutes the first experimental evidence of growing point-to-set correlations in glass-forming liquids. Further, $\xi_{dyn}$ grows faster than $\xi_{PTS}$, as observed in simulations \cite{kob2012non}. Most strikingly, however, $\xi_{dyn}$ exhibits a non-monotonic dependence on $\phi$ (Fig. \ref{Fig3}a). This result is remarkable, since it is the first experimental observation of non-monotonicity in dynamic correlations. With the exception of the numerical results of Kob et al. \cite{kob2012non}, all dynamic length scales reported in the literature were seen to grow monotonically on approaching the glass transition \cite{weeks2000three,kegel2000direct,berthier2005direct,flenner2012characterizing}. There are two potential causes for the paucity in observations of the aforementioned non-monotonicity. First, not all dynamic length scales are expected to show non-monotonicity and the presence of a pinned wall appears to be crucial to this observation. Even in the presence of a pinned wall, it has been shown that the existence of non-monotonicity is dependent on the interaction potential \cite{hocky2014crossovers}.  Secondly, the maximum in $\xi_{dyn}$ occurs at a characteristic temperature \cite{flenner2014universal,hocky2014crossovers} close to the mode coupling crossover, where sample equilibration is difficult even in simulations. An important point to note is that in the simulations of Kob and coworkers \cite{kob2012non} as well as our experiments, the maximum in $\xi_{dyn}$ occurs slightly before the mode coupling crossover (Fig. \ref{Fig3}a, Supplementary Fig. S1), strongly suggesting that our observations correspond to the same dynamic crossover seen in \cite{flenner2014universal,hocky2014crossovers}. In \cite{kob2012non}, it has been speculated that the observed non-monotonicty is a consequence of a change in the morphology of CRRs across the mode coupling crossover, and is therefore consistent with RFOT. In particular, the authors claim that the spatial inhomogeneity introduced by the wall makes $\xi_{dyn}$ sensitive not only to the number of particles in a CRR, but also to their arrangement into string-like or compact structures. 

To test whether the non-monotonicity indeed stems from a change in the shapes of CRRs, we examined the nature of dynamic heterogeneity in our system. We first identified the top 10$\%$ most mobile particles over various time intervals $\Delta t$ and clustered them based on nearest neighbour distances. As expected, the mean cluster size is maximal at a characteristic time $t^*$. On observing the shapes of these clusters defined over $\Delta t = t^*$, we find that quite remarkably, the clusters are predominantly string-like at low $\phi$ and compact at high $\phi$ (Fig. \ref{Fig3}b-c). To quantify this change in morphology, we computed the distribution $P(NN)$ of the number of mobile nearest neighbors of a mobile particle, following the protocol used in \cite{weeks2000three,zhang2011cooperative} (Fig. \ref{Fig3}d). Since small clusters might bias the distribution towards smaller values of $NN$, we only considered clusters having more than 4 particles. Interestingly, we observe that for $\phi < 0.76$, $P(NN)$ exhibits a peak at $NN=$ 2, indicating string-like morphology. For $\phi =0.76$, the distribution becomes broader, although the maximum remains at $NN=$ 2. Strikingly, for $\phi > 0.76$,  we observe a drastic change in the distribution with the peak shifting to $NN=$ 4, consistent with Fig. \ref{Fig3}b-c. Although we could sample only a limited number of $\phi$s owing to experimental difficulties, it is evident from the data that the maximum in $\xi_{dyn}$ (Fig. \ref{Fig3}a) coincides with the crossover in the morphology of CRRs (Fig. \ref{Fig3}d). These findings therefore provide direct confirmation that the non-monotonicity in $\xi_{dyn}$ results from a change in the shapes of CRRs. Further, we observe that the average number of particles per cluster increases monotonically (Supplementary Fig. S2). This is consistent with previous studies on colloidal glass-formers \cite{weeks2000three,kegel2000direct} and strongly suggests that length scales that grow monotonically on approaching the glass transition are sensitive only to the number of particles in a CRR and not to their arrangement within it. In a broader context, our results provide the first direct verification of the prediction of RFOT for the change in morphology of CRRs across the mode coupling crossover \cite{stevenson2006shapes}. 

Our results provide the first experimental evidence for growing amorphous order in glass forming liquids. In addition, we observe a non-monotonic dependence of $\xi_{dyn}$ on $\phi$ as well as a corresponding change in the morphology of CRRs, which indicates a crossover in the mechanism of relaxation from flow-like to activated dynamics. These observations also suggest that the recently observed non-monotonicity in the Mobility Transfer Function \cite{gokhale2014growing}, a measure of dynamical facilitation \cite{chandler2009dynamics}, may be due to the prominence of activated dynamics, rather than finite size effects \cite{elmatad2012manifestations}. On the other hand, while facilitation plays a subordinate role within RFOT \cite{bhattacharya2008facilitation}, recent experiments have shown that it significantly influences structural relaxation \cite{gokhale2014growing} and can even predict the existence of reentrant glass transitions \cite{mishra2014dynamical}. It is therefore imperative to examine the present results in the context of facilitation. Specifically, it would be fascinating to see whether the observed change in morphology of CRRs, which finds a natural explanation within RFOT, can also emerge within the facilitation picture. The influence of a pinned wall on facilitated dynamics is also worthy of exploration. Given that in the time averaged images (Fig. \ref{Fig1}A) we see signatures of increasing cluster size of immobile particles with $\phi$, it would be fascinating to explore connections between these regions of slow dynamics and the length scales extracted here. It would also be instructive to investigate whether our findings are consistent with other thermodynamic frameworks such as geometric frustration-based models \cite{tarjus2005frustration}. We expect our findings to engender future research aimed at addressing these unresolved issues on the theoretical, numerical as well as experimental front.

\section*{Materials and Methods}
Our system consisted of a binary mixture of $N_S$ small and $N_L$ large polystyrene particles of diameters $\sigma_S =$ 1.05 $\mu$m and $\sigma_L =$ 1.4 $\mu$m, respectively. The particle size ratio $\sigma_L/\sigma_S =$ 1.3 and number ratio $N_S/N_L =$ 1.23 provided sufficient geometric frustration to prevent crystallization. The samples were loaded into a wedge shaped cell \cite{gokhale2014growing} and the desired area fractions $\phi$ were attained by sedimentation of the sample to the monolayer-thick region of the wedge. Samples were imaged using a Leica DMI 6000B optical microscope with a 100X objective (Plan apochromat, NA 1.4, oil immersion) and images were captured at frame rates ranging from 3.3 fps to 5 fps for 1-3 hours depending on the values of $\phi$. The holographic optical tweezers set up consisted of a linearly polarized constant power (800 mW) CW laser (Spectra-Physics, $\lambda$ = 1064 nm). Standard Matlab algorithms \cite{crocker1996methods} were used to generate particle trajectories and subsequent analysis was performed using codes developed in-house. 
In order to set the reference for measuring z, we first calculated overlap functions for all the boxes lying within strips of width 0.5$\sigma_{s}$, parallel to the wall, i.e. along X-axis, for the entire image. From these overlap functions, we estimated the Y-coordinate for which the overlaps exhibit no decay and labelled that as the center of the wall. Given that the wall is $\sim$2 particle diameters wide, the overlap does not decay with time only for a few strips of 0.5 $\sigma_{s}$ away from the center of the wall. We have chosen the center of last strip for which the overlap function does not exhibit a decaying profile with time, as the reference from which to measure z.

\section*{Acknowledgements}
The authors thank Walter Kob for illuminating discussions. K.H.N. thanks the Council for Scientific and Industrial Research (CSIR), India for a Senior Research Fellowship. S.G. thanks CSIR, India for a Shyama Prasad Mukherjee Fellowship. A.K.S. thanks Department of Science and Technology (DST), India for support under J.C. Bose Fellowship and R.G. thanks the International Centre for Materials Science (ICMS) and the Jawaharlal Nehru Centre for Advanced Scientific Research (JNCASR) for financial support. 

\subsection*{Author Contributions}
K.H.N., S.G., A.K.S. and R.G designed research. K.H.N. performed research and analysed data. K.H.N. and S.G. wrote the paper with inputs from A.K.S. and R.G..
\newline
$^{\ast}$ Corresponding author

\newpage

\newpage

\newpage

\begin{figure}[tbp]
\begin{center}
\includegraphics[width=0.6\textwidth]{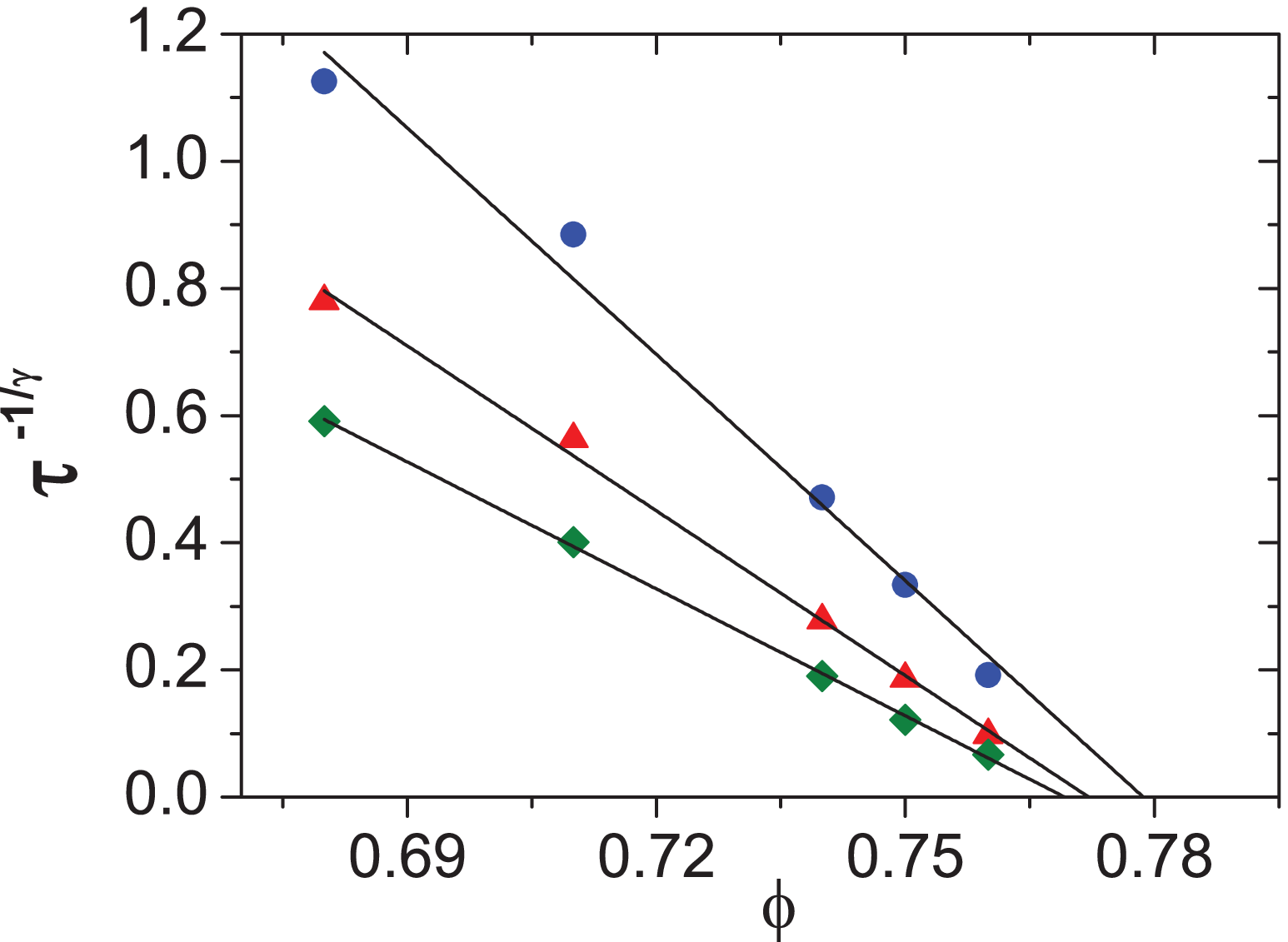}
\end{center}
\noindent \textbf{Fig. S1:} \textbf{Estimation of Mode coupling crossover $\phi$, $\phi_{MCT}$.} $\tau^{-\frac{1}{\gamma}}$ versus $\phi$ for wave vectors, $q$ = 2$\pi$/0.75$\sigma$ ({\color{Blue} $\boldsymbol \bullet$}), $q$ = 2$\pi $ / $\sigma$ ({\color{Red} $\boldsymbol \blacktriangle$}) and $q$ = 2$\pi$/1.25$\sigma$ ({\color{Green} $\boldsymbol \blacklozenge$}), where $\sigma$ is the mean diameter of the big and the small particles. Here, $\gamma$ = $\dfrac{1}{2a}+\dfrac{1}{2b} = 1.89$ has been computed by substituting $a=0.377$ obtained from ref \cite{gotze} and $b=0.885$ extracted from $F_{s}(q,t)$.  
\label{S1}
\end{figure}

\begin{figure}[tbp]
\begin{center}
\includegraphics[width=0.6\textwidth]{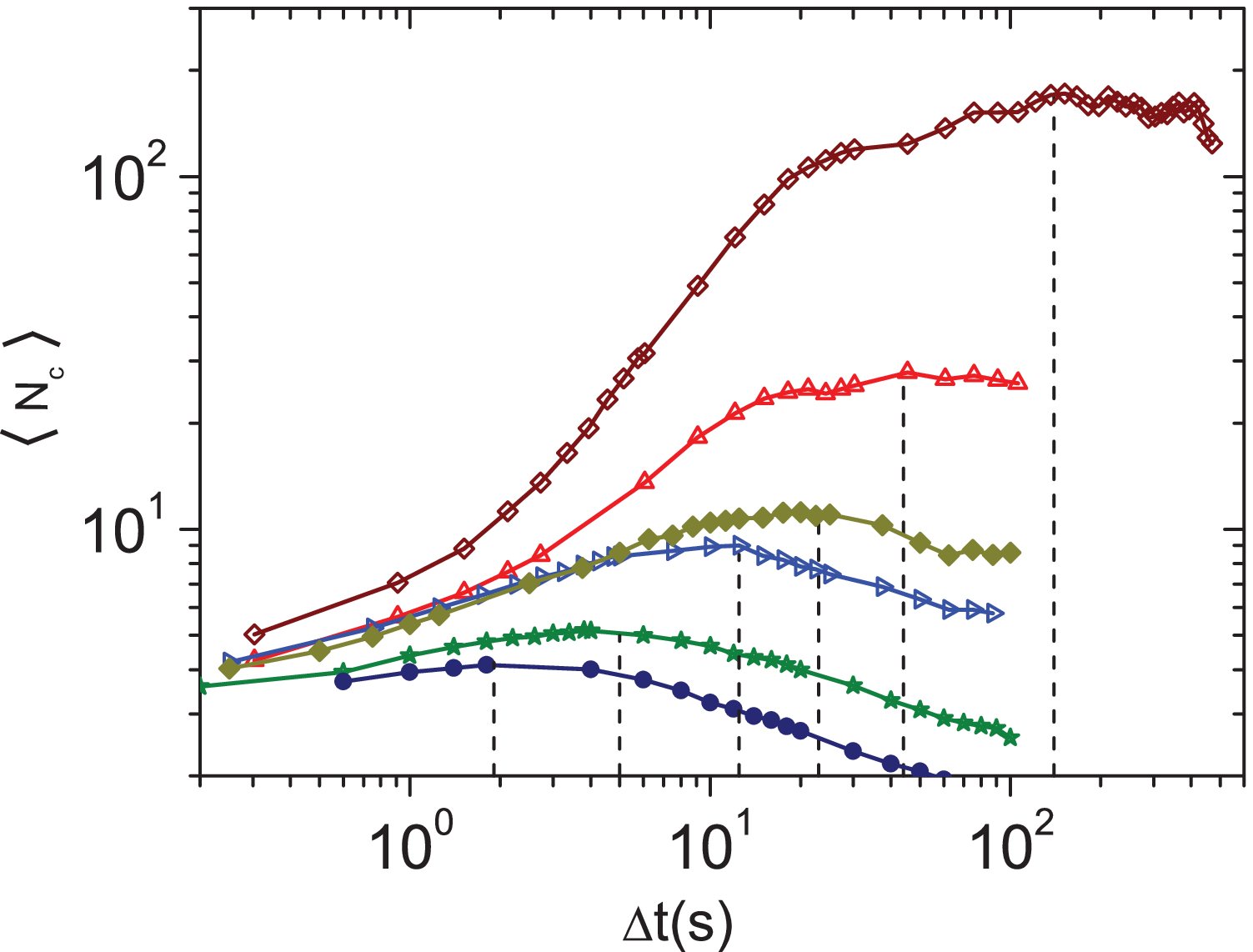}
\end{center}
\noindent \textbf{Fig. S2:} \textbf{Dependence of average cluster size on $\phi$.} Average cluster size, $\langle N_{c} \rangle$, as a function of $\Delta t$ for $\phi =$ 0.68 ({\color{Blue} $\boldsymbol \bullet$}), $\phi =$ 0.71 ({\color{Green} $\boldsymbol \star$}), $\phi =$ 0.74 ({\color{Blue} $\boldsymbol \triangleright$}), $\phi =$ 0.75 ({\color[rgb]{0.71,0.65,0.26} $\boldsymbol \blacklozenge$}), $\phi =$ 0.76 ({\color{Red} $\boldsymbol \blacktriangle$}) and $\phi =$ 0.79 ({\color{Brown} $\boldsymbol \diamond$}). The dotted lines correspond to the maximum in $\langle N_{c} \rangle$.
\label{S2}
\end{figure}

\end{document}